\documentclass[runningheads]{llncs}
\usepackage[T1]{fontenc}
\usepackage{multirow}
\usepackage{multicol}
\usepackage{booktabs}
\usepackage{array}
\usepackage{subcaption}
\usepackage{hyperref}       
\usepackage{xcolor}
\usepackage{sidecap}
\usepackage{amsmath}
\usepackage{amssymb}
\usepackage{mathtools}
\hypersetup{
    colorlinks,
    linkcolor={red!50!black},
    citecolor={blue!50!black},
    urlcolor={blue!50!black}
}
\usepackage{tabularray}
\usepackage{graphicx}

\makeatletter
\newcommand{\printfnsymbol}[1]{%
  \textsuperscript{\@fnsymbol{#1}}%
}
\makeatother

\usepackage{graphicx,verbatim}
\begin{document}
\title{Cardiovascular disease classification using radiomics and geometric features from cardiac CT}
\titlerunning{CVD classification with radiomic and geometric features}

\author{Ajay Mittal \inst{1}, Raghav Mehta \inst{1}, Omar Todd \inst{1}, \\ Philipp Seeböck \inst{2}, Georg Langs \inst{2}, Ben Glocker \inst{1} }
\institute{Imperial College London, UK. \\ \and Medical University of Vienna, Austria. \\
    \email{raghav.mehta@imperial.ac.uk}}

\authorrunning{A. Mittal, R. Mehta, O. Todd, P. Seeböck, G. Langs, B. Glocker}


\maketitle              
\begin{abstract}
Automatic detection and classification of Cardiovascular disease (CVD) from Computed Tomography (CT) images play an important part in facilitating better-informed clinical decisions. However, most of the recent deep learning based methods either directly work on raw CT data or utilize it in pair with anatomical cardiac structure segmentation by training an end-to-end classifier. As such, these approaches become much more difficult to interpret from a clinical perspective. To address this challenge, in this work, we break down the CVD classification pipeline into three components: (i) image segmentation, (ii) image registration, and (iii) downstream CVD classification. Specifically, we utilize the Atlas-ISTN framework and recent segmentation foundational models to generate anatomical structure segmentation and a normative healthy atlas. These are further utilized to extract clinically interpretable radiomic features as well as deformation field based geometric features (through atlas registration) for CVD classification. Our experiments on the publicly available ASOCA dataset show that utilizing these features leads to better CVD classification accuracy (87.50\%) when compared against classification model trained directly on raw CT images (67.50\%). Our code is publicly available: \url{https://github.com/biomedia-mira/grc-net}

\keywords{Cardiovascular Disease \and CT Image \and Foundational Models.}

\end{abstract}

\section{Introduction}
Cardiovascular diseases (CVDs) remain the leading cause of morbidity and mortality worldwide, accounting for millions of deaths each year. In Europe, more than 11 million new cases of CVD occur each year \cite{wilkins2017european}. As such, early and accurate diagnosis is critical to improving patient outcomes and reducing the burden on healthcare providers. Computed Tomography (CT) has emerged as a powerful non-invasive tool for visualizing the cardiovascular system, providing detailed clear anatomical and functional information. However, manual interpretation of CT images is time-consuming, prone to inter-observer variability, and often requires specialized expertise. This underscores an urgent need for automated systems that can assist clinicians in detecting CVD with high accuracy. 

Recent advances in machine learning models, specifically deep learning models, have allowed better automatic cardiac image analysis \cite{martin2020image}. These methods have become integral part of cardiovascular disease (CVD) classification from CT imaging, offering fast and accurate alternatives to manual interpretation \cite{bray2022machine}. Convolutional neural networks (CNNs) have been widely used to detect and quantify coronary artery disease from coronary CT angiography and calcium scoring CT \cite{lin2022deep,thanassoulis2012genetic}, showing high agreement with expert annotations and prognostic value. More recent approaches have applied vision transformers \cite{alven2025plaquevit} to segment vessels and classify plaque burden with improved generalization. Deep learning has also been used to assess valvular diseases, such as aortic stenosis, by quantifying calcification from cardiac CT \cite{park2025deep}. For heart failure evaluation, CNNs enable precise segmentation of cardiac chambers, supporting functional assessment from volumetric CT data \cite{bruns2022deep}. Multi-modal models that integrate CT-derived features with clinical variables are also emerging to enhance diagnostic accuracy \cite{amal2022use}. These advancements demonstrate the growing effectiveness of deep learning in improving CVD diagnosis across various CT modalities.

Most of the aforementioned deep learning-based approaches for CVD classification either operate directly on raw CT images or utilize it with structural segmentation masks for disease classification. However, these methods are not clinically interpretable. In addition to that, segmentation-based features do not capture information about the geometry of different anatomical structures. In this paper, we hypothesize that geometric features related to the morphology of anatomical structures (ex. deformation field to healthy atlas)  also contain important clinical and diagnostic information. These geometric features provide complementary insights to those derived from structural segmentation. Therefore, leveraging both types of features in conjunction could lead to improved CVD classification performance and interpretability. Our experiments and results on the publicly available ASOCA dataset \cite{gharleghi2022automated} support this hypothesis and demonstrate a 5\% improvement in classification performance.

\section{Methodology}

\begin{figure}[t]
    \centering
    \includegraphics[width=1\linewidth]{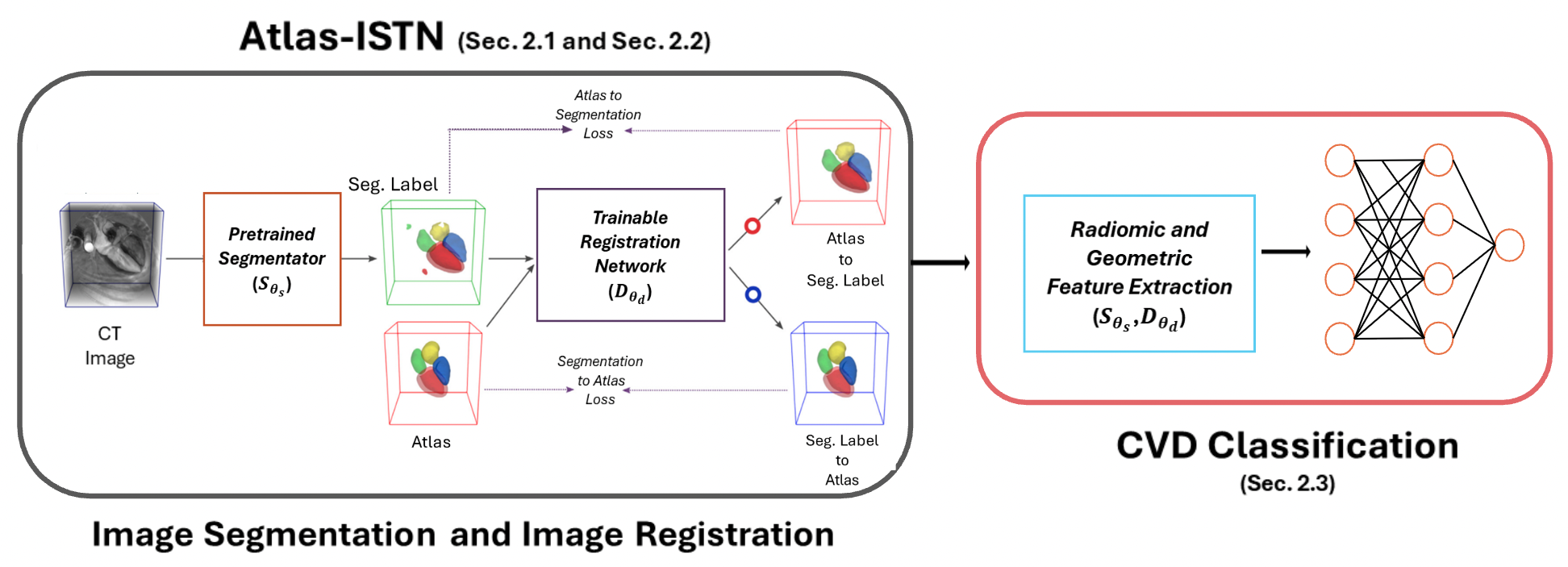}
    \caption{Overview of the proposed multi-stage CVD classification framework.}
    \label{fig:over}
\end{figure}

As shown in Fig.\ref{fig:over}, we divide the CVD classification pipeline into three components: (i) image segmentation, (ii) image registration to population atlas, and (iii) downstream CVD classification. For image segmentation and registration, we employ Atlas-ISTN \cite{sinclair2022atlas}, which has shown to produce clinically meaningful large structure segmentation in relation to a population atlas. Furthermore, we exploit pre-trained foundational models within the Atlas-ISTN framework to improve performance in the low-data regime. Using the generated structure segmentation and atlas registration, we extract radiomic and geometric features. These features are utilized for the CVD classification, improving performance and interpretability. 

\subsection{Background on Atlas-ISTN} \label{sec:AISTN}
Deep learning models for cardiac structure segmentation learn robust voxel-wise features but are often sensitive to test-time noise, leading to anatomically implausible predictions. Conversely, image registration models preserve anatomical topology through spatial transformations but typically require extensive training data. The Atlas Image-and-Spatial Transformer Network (Atlas-ISTN \cite{sinclair2022atlas}) addresses these challenges by jointly learning segmentation and registration while constructing a population-based anatomical atlas. This atlas captures statistical anatomical variation and enables extraction of geometric features such as deformation fields, which quantify deviations from the normative healthy anatomy and can aid in abnormality detection. To mitigate the reliance on large labeled datasets for segmentation, in this work, we leverage pretrained foundational models like TotalSegmentator \cite{wasserthal2023totalsegmentator} and Anatomix \cite{dey2024learning}, within Atlas-ISTN.  

\subsection{Background on Segmentation Foundational Models}

TotalSegmentator \cite{wasserthal2023totalsegmentator} is a pre-trained, foundational segmentation model capable of generating segmentations for up to 117 different anatomical structures. Its architecture is based on nnU-Net \cite{isensee2021nnu}, a well-established benchmark in medical image segmentation. TotalSegmentator was pre-trained to segment major cardiac structures using a publicly available cardiac CT angiography (CCTA) dataset ($n \approx 1200$). This enables TotalSegmentator to segment the following seven major structures for cardiac CT image analysis: right atrium, left atrium, right ventricle, left ventricle, left ventricular myocardium, aorta, and pulmonary trunk. Notably, since the model is specifically designed for CCTA images, it can be directly utilized to generate structural segmentations and seamlessly integrated into the Atlas-ISTN framework without any modifications or fine-tuning.

Anatomix \cite{dey2024learning} is a general-purpose foundational model developed to support a variety of medical image analysis tasks, including segmentation. It is built upon a 3D U-Net \cite{cciccek20163d} architecture and pre-trained on large volumes of synthetic anatomical data, allowing for robust generalization across diverse modalities and anatomical structures without the need for retraining from scratch. A key strength of Anatomix lies in its ability to leverage this pre-trained network for domain-specific segmentation using only a limited amount of annotated data. Consequently, Anatomix can be effectively integrated into the Atlas-ISTN framework by fine-tuning it on a small set of images.

\subsection{Disease classification}

\begin{figure}[t]
    \centering
    \includegraphics[width=1.0\linewidth]{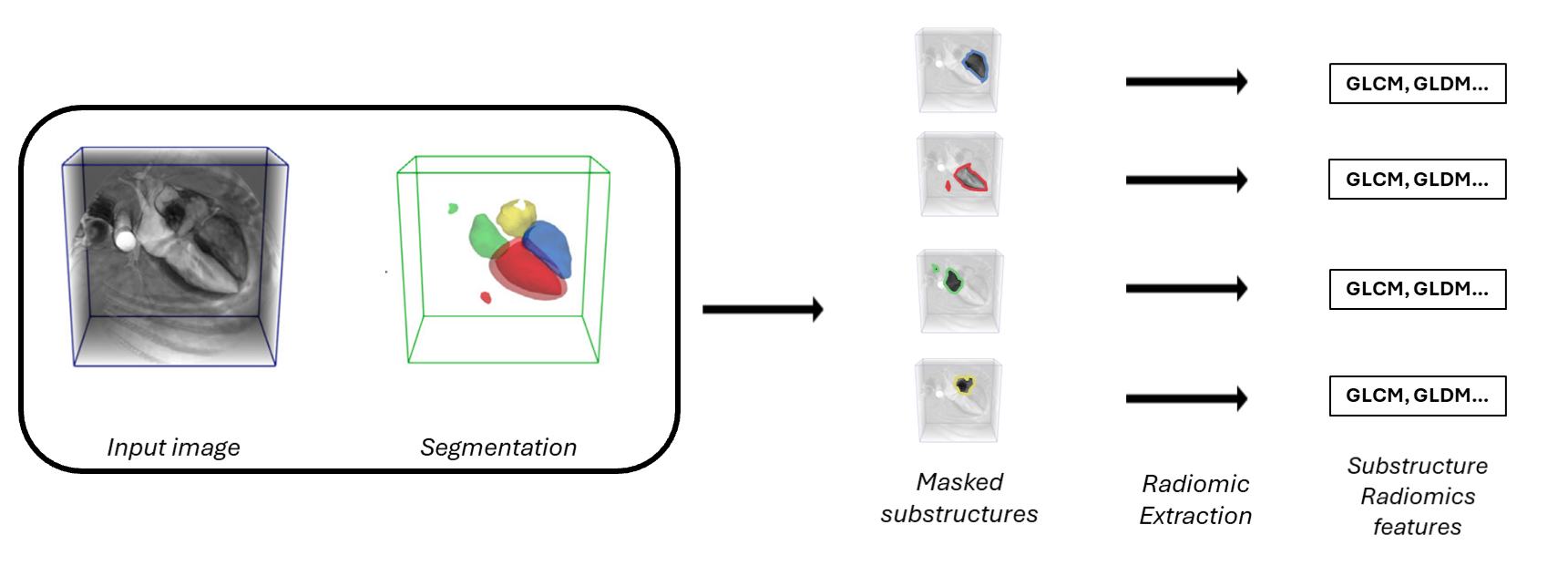}
    \caption{Pipeline for radiomic feature extraction per image. The input image and corresponding segmentation are used to isolate individual anatomical substructures. Each masked region is then processed to extract radiomic features, producing per-structure features summarising shape and intensity characteristics.}
    \label{fig:RAD}
\end{figure}

\begin{figure}[t]
    \centering
    \includegraphics[width=1\linewidth]{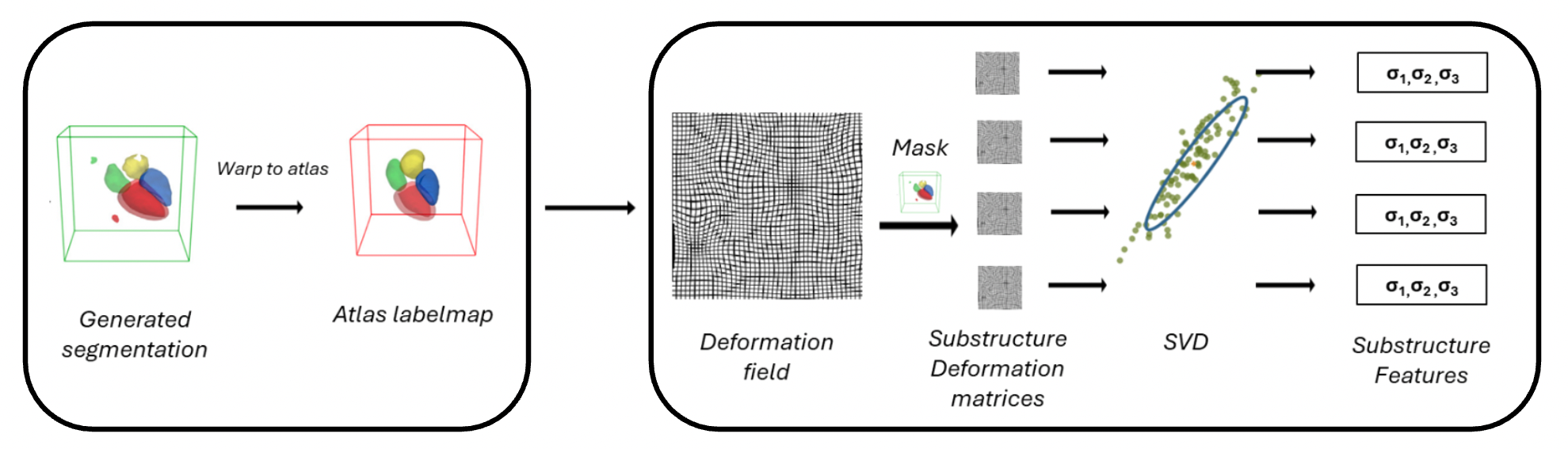}
    \caption{Pipeline for computing interpretable deformation features. Left: Predicted labelmap of an image (green box) and the healthy cardiac atlas (red box). Center-left: The deformation field estimated by registering the subject labelmap to the atlas. Center-right: The deformation field is masked per anatomical structure to isolate substructure-specific displacements. Right: These displacements are decomposed using SVD to extract low-dimensional, interpretable features.}
    \label{fig:SVD}
\end{figure}

We utilize the generated segmentation and atlas from the Atlas-ISTN framework for the downstream CVD classification task. Specifically, we extract radiomic features \cite{van2017computational} from the generated segmentation maps, and geometric features from the deformation fields produced by warping each image to the constructed atlas.

Radiomic features \cite{van2017computational}, originally developed in the context of cancer research, represent a class of standardized, quantifiable information of lower dimensionality extracted from delineated regions of medical images. We extract the following features (Fig. \ref{fig:RAD}) from each of the seven segmented cardiac structures:

\begin{itemize}
\item First-order statistics describes the distribution of voxel intensities, including metrics such as mean, variance, skewness, kurtosis, and entropy.
\item Shape descriptors (3D) characterizes the geometry of each substructure, encompassing attributes such as volume, surface area, compactness, sphericity, and elongation.
\item GLCM (Gray Level Co-occurrence Matrix) quantifies spatial relationships between pairs of voxel intensities, capturing properties such as contrast, correlation, and homogeneity.
\item GLRLM (Gray Level Run Length Matrix) measures the length of consecutive voxels with the same intensity, indicating texture coarseness or streaking.
\item GLSZM (Gray Level Size Zone Matrix) assesses the size of homogeneous regions, independent of orientation, to capture the granularity of structures.
\item NGTDM (Neighbouring Gray Tone Difference Matrix) captures local texture by evaluating intensity differences between a voxel and its neighboring voxels.
\item GLDM (Gray Level Dependence Matrix) measures the number of connected voxels within a specific intensity range, providing texture complexity.
\end{itemize}

To extract geometric features, we compute voxel-wise displacement vectors from each image to the normative healthy atlas. Since deformation fields are spatially dense and high-dimensional, we apply dimensionality reduction techniques, such as Singular Value Decomposition (SVD), to reduce their feature space. Similar to radiomic features, this process is performed individually for each anatomical structure (Fig. \ref{fig:SVD}).

\section{Experiments and Results}

\subsection{Datasets}
We utilize two distinct cardiac CT datasets: one for the segmentation task and the other for the downstream disease classification task. Specifically, we use the Multi-Modality Whole Heart Segmentation (MM-WHS) dataset \cite{zhuang2019evaluation}, which is widely recognized as a benchmark for cardiac segmentation tasks. This dataset includes 20 labeled CT and 40 unlabeled MRI images. For the purpose of this study, we focus exclusively on the CT images, consisting of 20 high-resolution 3D cardiac volumes. Each labeled volume includes annotations for seven cardiac substructures: the left and right ventricles, left and right atria, myocardium, descending aorta, and pulmonary artery. In our experiments, we split the CT dataset into training, validation, and testing sets in a 13/1/6 ratio, respectively. For the disease classification task, we utilize the ASOCA dataset \cite{gharleghi2022automated}, a publicly available 3D cardiac CT dataset comprising 20 healthy and 20 diseased cases. Healthy cases are defined by a zero calcium score and no evidence of vascular disease, while diseased cases exhibit either significant calcification or stenosis as diagnosed by a cardiologist. Due to the limited size of the dataset, we employ 5-fold cross-validation for evaluation.

\subsection{Implementation Details}

Recall that TotalSegmentator is a pretrained model that does not require fine-tuning for cardiac CT segmentation of the same seven structures; in contrast to that, Anatomix requires fine-tuning. Accordingly, we trained Anatomix on the MM-WHS dataset using the dataset split described above. For comparison, we also trained a 3D U-Net model from scratch on the same split, serving as a baseline. After training the segmentation models and freezing their weights, the Atlas-ISTN model was trained using the same dataset to generate healthy atlas.

For the disease classification task, we employ a simple multilayer perceptron (MLP). To disentangle the contributions of radiomic and deformation-based geometric features, we trained three separate MLP models: (i) only radiomic features, (ii) only geometric features, and (iii) using a combination of both. The following hyperparameters were tuned using Bayesian Optimization \cite{akiba2019optuna} for each feature set: Hidden layer count \(L\in[1,12]\), Hidden unit size \(H\in[8,512]\), Dropout probability \(p\in[0,0.5]\), Learning rate \(\eta\in[10^{-4},10^{-2}]\) (log-uniform), Number of training epochs \(E\in\{100,125,150,\ldots,400\}\), Number of retained SVD singular values for deformation per structure \(\mathrm{n\_svd}\in[1,3]\). As a baseline, we also trained a 3D ResNet-50 \cite{feichtenhofer2018slowfast} classifier directly on the CT images. All models were trained using the AdamW optimizer \cite{loshchilov2017decoupled} and binary cross entropy loss.

\begin{figure}[t]
    \centering
    \includegraphics[width=1\linewidth]{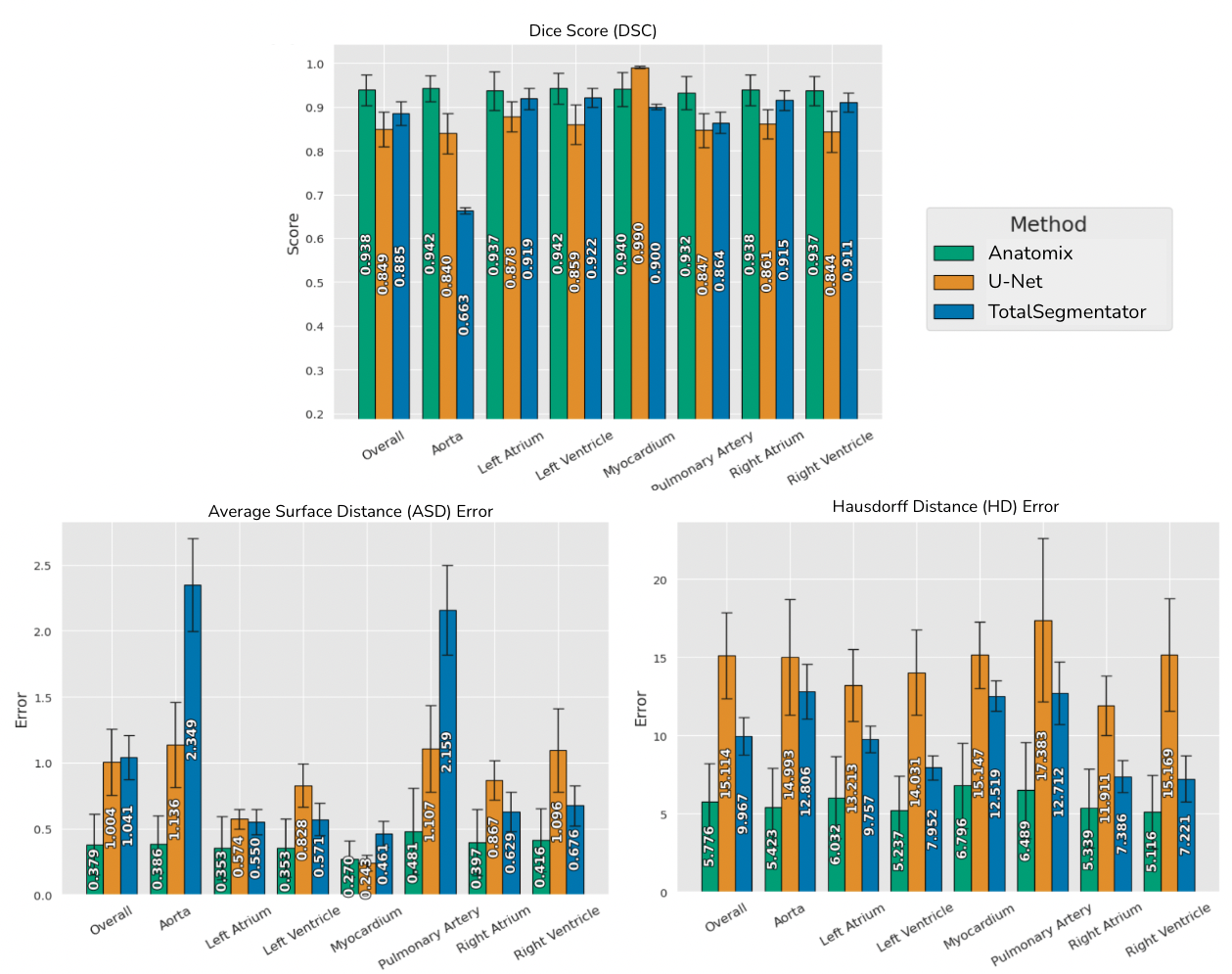}
    \caption{Segmentation performance metrics for each segmentation model.}
    \label{fig:quants}
\end{figure}

\subsection{Results: Image Segmentation and Atlas-Construction}

We evaluate segmentation performance using standard metrics such as Dice Similarity Coefficient (DSC), Hausdorff Distance (HD), and Average Surface Distance (ASD). For each metric, we report both global values (mean across all structures) and per-substructure values.

We can see in Fig.\ref{fig:quants} that both Anatomix and TotalSegmentator achieve high overlap (DSC) with the manual labels, demonstrating that modern pretrained or fine-tunable models can produce segmentation results closely aligned with ground truth, even with limited training data. In contrast, the 3D U-Net underperforms, suggesting that a fully supervised network trained solely on the MM-WHS subset struggles to generalize across all cardiac structures. Notably, for TotalSegmentator, the Dice score for the aorta is considerably lower than that of the other models. This reduced performance is likely due to over-segmentation of the aorta, where regions such as the descending aorta, excluded from the ground truth, are included, thereby reducing overall overlap scores. When considering the ASD and HD, Anatomix consistently achieves the lowest boundary errors, indicating superior accuracy in delineating anatomical surfaces. TotalSegmentator exhibits slightly higher boundary errors, again with the aorta as an outlier. The 3D U-Net, however, shows the highest boundary errors overall, further emphasizing that its limited supervised training fails to adequately preserve anatomical boundaries.

\begin{figure}[t]
    \centering
    \includegraphics[width=1\linewidth]{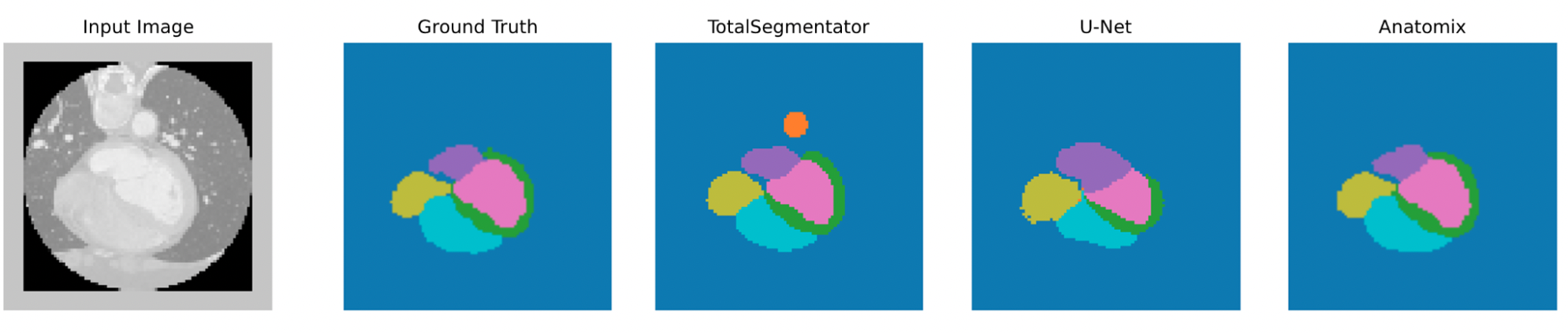}
    \includegraphics[width=1\linewidth]{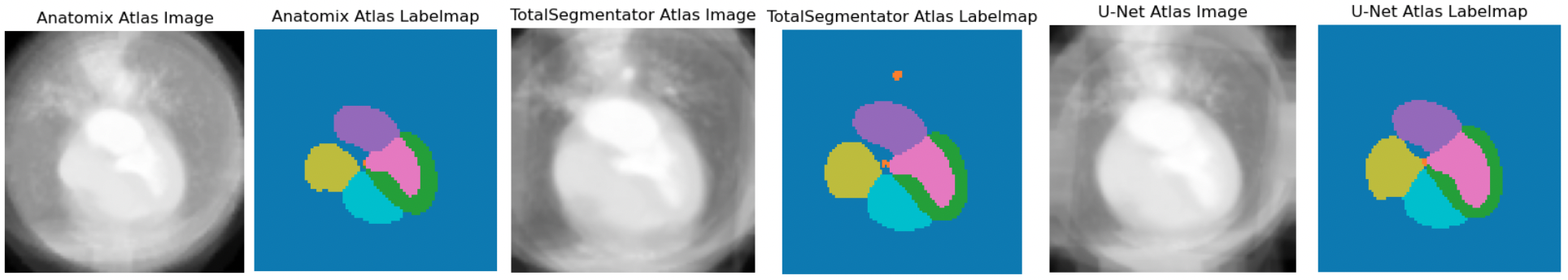}
    \caption{Comparison of (top-row) segmentation maps and (bottom-row) atlases generated by different methods.}
    \label{fig:enter-label-quals}
\end{figure}

While quantitative metrics summarize model performance, qualitative analysis provides deeper insights. Fig.\ref{fig:enter-label-quals}(top-row) compares model-generated segmentations with input images and ground-truth annotations. TotalSegmentator’s output is smooth and anatomically plausible, reflecting strong training priors. Though generally aligned with the ground truth, it includes the descending aorta, unlabeled in the MM-WHS dataset, leading to over-segmentation above the heart and explaining its lower Dice scores. In contrast, U-Net predictions show artifacts; the right atrium appears fragmented, and overall shapes are less refined. Anatomix produces the most visually consistent segmentation, closely matching the ground truth and omitting the descending aorta, likely due to fine-tuning on MM-WHS. Fig.\ref{fig:enter-label-quals}(bottom-row) shows the image and label atlases from different methods. U-Net-trained Atlas-ISTN yields the blurriest image atlas, while those from Anatomix and TotalSegmentator are sharper. However, the TotalSegmentator atlas contains spurious labels, unlike the cleaner Anatomix atlas.

\subsection{Results: Disease Classification}

\begin{table}[t]
\centering
\caption{Disease classification performance using different feature sets and model types. Values are reported as mean ± standard deviation over 5-fold cross-validation and 3 random seeds. The best-performing method is shown in bold.}
\resizebox{\textwidth}{!}{
\begin{tabular}{lcccc}
\toprule
\textbf{Metric}       & \textbf{Radiomic + Geometric} & \textbf{Radiomic Only} & \textbf{Geometric Only} & \textbf{ResNet-50 Baseline} \\
\midrule
Accuracy (\%)              & \textbf{87.50 \tiny{± 10.21}} & 82.50 \tiny{± 11.50} & 76.67 \tiny{± 11.96} & 67.50 \tiny{± 06.12} \\
Precision (\%)            & \textbf{88.11 \tiny{± 13.70}} & 83.60 \tiny{± 14.80} & 83.56 \tiny{± 16.67} & 84.76 \tiny{± 18.90}\\
Recall (\%)               & \textbf{90.00 \tiny{± 12.25}} & 85.00 \tiny{± 14.00} & 73.33 \tiny{± 23.21} & 60.00 \tiny{± 33.91} \\
F1 Score (\%)             & \textbf{88.01 \tiny{± 09.27}} & 83.70 \tiny{± 10.80} & 74.35 \tiny{± 15.08} & 59.88 \tiny{± 16.77} \\
Specificity (\%)          & \textbf{85.00 \tiny{± 17.80}} & 80.00 \tiny{± 19.00} & 80.00 \tiny{± 20.82} & 75.00 \tiny{± 31.62} \\
\bottomrule
\end{tabular}
}
\label{tab:disease_classification_results}
\end{table}

We evaluate disease classification performance using standard metrics such as accuracy, precision, recall, F1 score, and specificity. For feature extraction, we utilized segmentations generated by Anatomix, as it demonstrated the best performance among competing methods in the previous section.

The results reported in Table~\ref{tab:disease_classification_results} indicate that combining radiomic and deformation features yields the highest overall performance. The MLP classifier achieved an average accuracy of 87.50\%, with balanced precision (88.11\%), recall (90\%), and F1 score (88.01\%). Models using only radiomic or only deformation features performed worse individually, with accuracy dropping by approximately 5\% and 11\%, respectively, suggesting that these two feature types provide complementary information. The ResNet-50 baseline, trained directly on images, underperformed in both accuracy (67.5\%) and recall (60\%), despite exhibiting high precision, indicating poor generalization to positive cases. These results support the hypothesis that interpretable features derived from anatomical segmentations and deformation fields are more discriminative than features learned directly from raw intensity data, and disease classification performance can be improved without reliance on large image datasets or end-to-end training.

\section{Conclusion} 

In this paper, we proposed enhancing the clinical interpretability of cardiovascular disease (CVD) classification by converting it into a multi-stage pipeline. Specifically, for CVD classification, we leverage large-structure segmentation-derived radiomic features alongside geometric features obtained through image registration to a normative healthy atlas. We demonstrated the utility of pretrained foundational models for accurate structure segmentation. Our experiments on the publicly available ASOCA dataset confirmed that radiomic and geometric features provide complementary information, with their combination yielding the best classification performance. As a future work, we plan to validate our findings on a much larger CVD classification dataset. In addition to that we will also incorporate uncertainty estimates associated with structure segmentation into the pipeline \cite{wundram2024leveraging,mehta2021propagating}, which could further improve both the interpretability and clinical utility of CVD classification.

\section*{Acknowledgment}
This project has received funding from the European Union’s Horizon Europe research and innovation programme under grant agreement 101080302 (AI-POD). 

\subsection*{Disclosure of interests}
B.G. is part-time employee of DeepHealth. No other competing interests.

\bibliography{References}

\end{document}